\documentclass[11pt]{article}
\usepackage[utf8]{inputenc}
\usepackage{fullpage}
\usepackage{graphicx}
\usepackage{amsmath}

\title{How Much Structure Is Needed for Huge Quantum Speedups?}
\author{Scott Aaronson\thanks{University of Texas at Austin. \ Email:
aaronson@cs.utexas.edu. \ Supported by a Vannevar Bush Fellowship from the US
Department of Defense, the Berkeley NSF-QLCI CIQC Center, a Simons Investigator Award, and the Simons
\textquotedblleft It from Qubit\textquotedblright\ collaboration.}}
\date{September 2022}

\begin{document}

\maketitle

\begin{abstract}
I survey, for a general scientific audience, three decades of research into
which sorts of problems admit exponential speedups via quantum
computers---from the classics (like the algorithms of Simon and
Shor), to the breakthrough of Yamakawa and
Zhandry from April 2022. \ I discuss both the quantum circuit model,
which is what we ultimately care about in practice but where our
knowledge is radically incomplete, and the so-called \emph{oracle} or
\emph{black-box} or \emph{query complexity} model, where we've managed to
achieve a much more thorough understanding that then informs our
conjectures about the circuit model. \ I discuss the strengths and
weaknesses of switching attention to sampling tasks, as was done in
the recent quantum supremacy experiments. \ I make some
skeptical remarks about widely-repeated claims of exponential quantum
speedups for practical machine learning and optimization problems. \
Through many examples, I try to convey the ``law of conservation
of weirdness,'' according to which every problem admitting an exponential
quantum speedup must have some unusual property to allow the amplitude
to be concentrated on the unknown right answer(s).

Edited transcript of a rapporteur talk delivered at the $28^{th}$ Solvay Physics Conference in Brussels, Belgium on May 21, 2022.
\end{abstract}

\section{Introduction}

I can't think of any better way to celebrate my $41^{st}$ birthday than to kick off the Solvay discussion of quantum algorithms! \ This talk \emph{will} include the birthday paradox, by the way.

How much structure does a problem have to have before the problem admits, let's say, an exponential speedup via a quantum computer? \ I'd submit that this is the some sense the central question of quantum algorithms research, from the beginnings of the subject in the early 1990s all the way till today.

\begin{figure}
\includegraphics[width=6.5in]{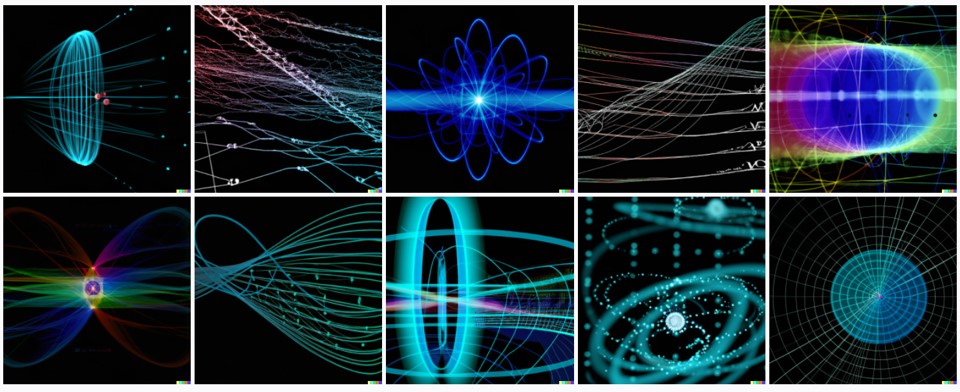}
\caption{``How Much Structure Is Needed for Huge Quantum Speedups?,'' according to DALL-E2}
\label{dalle}
\end{figure}

It's a difficult question. \ It's so difficult that I posed it to an AI artist, DALL-E2, getting back the striking images shown in Figure \ref{dalle}. \ There \emph{might} be some insight in them, I'm not sure.

My starting point is: to many of the companies, venture capitalists, government agencies, and so on now making investments in the field, quantum computing basically means magic. \ It means a pixie dust that lets you try all the possible solutions to your problem in parallel and then instantly pick the best one. \ Clearly such a pixie dust, if it existed, \emph{would} be a next logical step for computing after Moore's Law petered out. \ It could be just as transformative for civilization as classical computing has been.

Alas, we've understood for 30 years that the reality of quantum algorithms is subtler than that. \ I like to say that quantum computing is strange enough that no science fiction writer would've had the imagination to invent it.

\begin{figure}
\includegraphics[width=6.5in]{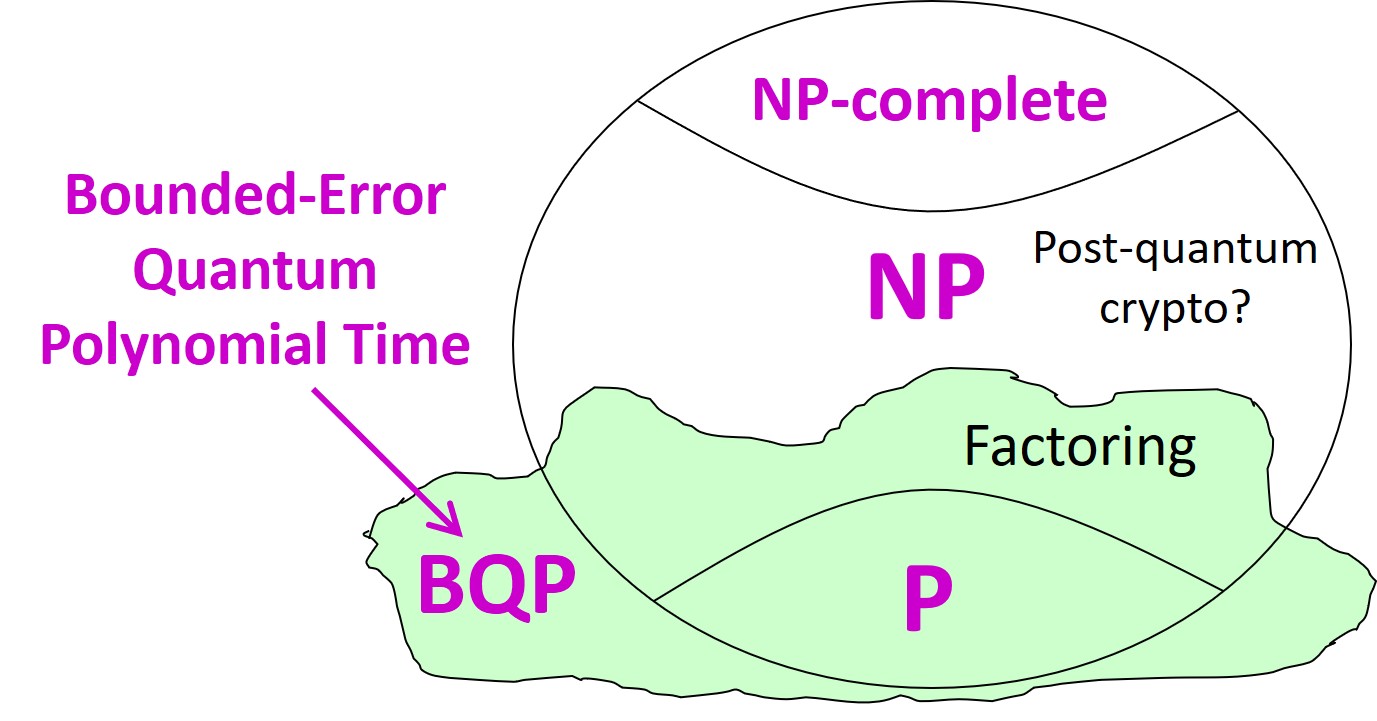}
\caption{Complexity classes relevant to this talk.}
\label{classes}
\end{figure}

So, where does quantum computing fit in to the landscape of computational complexity theory? \ John Preskill already sketched the picture in his beautiful opening talk for this conference \cite{preskill:solvay}, but Figure \ref{classes} shows what it looks like when drawn.

At the bottom we have $\mathsf{P}$, or Polynomial-Time, which is informally the class of all problems that are efficiently solvable by a classical computer.\footnote{If you care about the more detailed definition, more likely than not you already know it!} \ Above that we have $\mathsf{NP}$, Nondeterministic Polynomial-Time, the efficiently verifiable problems. \ The $\mathsf{NP}$-complete problems at the top are the hardest problems in $\mathsf{NP}$, meaning those to which all other $\mathsf{NP}$ problems can be efficiently reduced. \ And then we've got $\mathsf{BQP}$, Bounded-Error Quantum Polynomial-Time, which I drew with a wavy border because everything quantum is spooky and weird.

$\mathsf{BQP}$, indeed, fits a bit oddly into the structure. \ Peter Shor's great discovery \cite{shor} was that $\mathsf{BQP}$ contains certain specific $\mathsf{NP}$ problems, such as factoring integers and discrete logarithms, that are not known to be in $\mathsf{P}$. \ And these problems, as we all know, have enormous importance for modern cryptography. \ But they're also strongly believed \emph{not} to be $\mathsf{NP}$-complete. \ In order to give efficient quantum algorithms for these problems, Shor had to exploit special structure in them, involving periodicity, or (more abstractly) hidden subgroups of abelian groups. \ This structure sharply \emph{differentiates} factoring and discrete log from generic $\mathsf{NP}$ search problems, or even generic problems of breaking cryptographic codes.

So, one could ask: can we say anything insightful about the set of problems that admit huge quantum speedups? \ Say, the problems that are in $\mathsf{BQP}$ but not in $\mathsf{P}$?

I should emphasize that, in this talk, I won't be focused on polynomial speedups---as epitomized by the famous square-root speedup for unordered search, from Grover's algorithm \cite{grover}. \ As interesting as polynomial quantum speedups are to me as a theorist, the vast majority of known such speedups are quadratic or less. \ And recent estimates (see, e.g., \cite{babbush:beyondquadratic}) have made clear that, given the huge overheads required for quantum error-correction, it will be a long, \emph{long} time (if ever) before a quantum algorithm with a quadratic speedup would yield a net win over classical computing in practice. \ The practical outlook for \emph{exponential} quantum speedups, like that from Shor's algorithm, looks much more favorable.

With that out of the way, here's what I call the ``Hammer of Hype'': in order to be in the set $\mathsf{BQP}\setminus \mathsf{P}$, a problem needs to satisfy two conditions.

Condition 1---I hope you're sitting down for this!---is that the problem has to be in $\mathsf{BQP}$. \ In particular, you have to be able to choreograph a pattern constructive and destructive interference that concentrates a lot of amplitude on the right answer. \ You have to do that even though you yourself don't know in advance which answer \emph{is} the right one. \ Otherwise, what would be the point?

Condition 2 is that the problem has to not be in $\mathsf{P}$. \ So, the quantum algorithm has to beat the best classical algorithm that anyone can think of, by a superpolynomial factor. \ And classical computing gets to fight back!

This is a tall order. \ A priori, there didn't have to be \emph{any} interesting problems that satisfied both of the above conditions---or at least, any problems not about simulating quantum mechanics itself. \ It was one of great discoveries of my lifetime that there seemingly \emph{are} such problems, at least in the domains of number theory, group theory, and cryptography. \ But in the 1990s, such problems seemed pretty specialized---and while we've learned a lot more, they still seem pretty specialized today.

So then, what superpolynomial speedups do we believe to be real, after 30 years of quantum algorithms research?

\section{The Circuit Model and the Black-Box Model}

Before we jump in, a crucial thing to understand is that there are two very different kinds of knowledge about quantum speedups.

The first kind of knowledge concerns the circuit model of quantum computing. \ Here the resource that we're trying to minimize is simply the number of 1- and 2-qubit gates that the quantum computer has to apply. \ This is what we actually care about in practice: what can we do using a number of elementary operations that grows only polynomially with $n$, the size of the problem instance that we're trying to solve?

The trouble is that, while we know various nontrivial quantum algorithms---examples of things that we can do with small numbers of gates---we almost never have any understanding of whether those algorithms are optimal. \ So in particular, we can almost never prove reasonable \emph{lower} bounds on the number of gates that are needed to solve some concrete problem.

I hasten to add: this is not a failure of \emph{quantum} complexity theorists specifically! \ Instead, it's a consequence of complexity theory's more general inability to prove almost any strong lower bounds on circuit size, including \emph{classical} circuit size, unconditionally. \ If we knew how to prove, for example, that $\mathsf{NP}$-complete problems lack polynomial-size quantum circuits, then at a bare minimum we'd know how to prove $\mathsf{P} \neq \mathsf{NP}$---one of the great unproved conjectures of modern mathematics.

\begin{figure}
\centering
\includegraphics[width=4in]{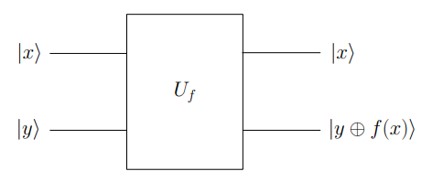}
\caption{The unitary transformation $U_f$ that corresponds to querying a function $f$.}
\label{query}
\end{figure}

But that brings me to our second kind of knowledge about quantum speedups. \ A large fraction of what we know about quantum algorithms is in what's called the black box or oracle or query complexity model (those terms all mean the same thing). \ Here the resource to be minimized is the number of queries or accesses to a black box, which reversibly computes a certain function $f$. \ Figure \ref{query} shows what it looks like as a unitary transformation.

Now, since this is quantum computing, you're allowed to call the black box on a superposition of inputs. \ But you can learn about the function $f$ \emph{only} by calling the black box.

The disadvantage of this model is that it's relevant to real life only under the assumptions that:

\begin{enumerate}
\item[(1)] you know how to apply this operation $U_f$ (that is, how to compute $f$), and
\item[(2)] you \emph{don't} know the internals of how the algorithm to compute $f$ works. \ (For if you knew those internals, you might be able to exploit them to go outside the black-box model.)
\end{enumerate}

The compensating advantage is that, in the black-box model, a detailed understanding is often achievable. \ With a lot of work, we've often been able to prove matching upper and lower bounds on the number of queries needed by a quantum computer to solve a given problem.

Maybe I could explain this to physicists by saying that the black-box model is theoretical computer science's Euclidean path integral, or our perturbation theory. \ It's something that we work on, not because it always tells us what we want to know, but because we can do the calculation.

But in addition, the black-box model has a long history of giving us insight about the ``real'' world---the circuit world. \ We heard a great example of that a few days ago, when Peter Shor told us how Simon's algorithm \cite{simon}, which is a black-box algorithm, gave him the crucial insight that he needed to invent his quantum algorithms for factoring and discrete logarithms (see \cite{shor:solvay}). \ As a much smaller example, Alex Arkhipov and I also invented BosonSampling \cite{aark} by starting with a black-box problem, and then looking for ways to remove the black box.

\section{A Tour of Exponential Quantum Speedups}

But let me now discuss the question: what exponential quantum speedups do we currently feel confident about in the ``real'' world---that is, the circuit world?

Obviously, there's factoring and discrete log. \ And beyond those, there's a whole slew of problems that involve calculating the order, or various aspects of the structure, of abelian groups (usually finite, sometimes infinite)---see, e.g., \cite{hallgren}. \ We can extend this \emph{slightly} into the realm of nonabelian groups, although progress has largely stalled there (see e.g.\ \cite{hmrrs} for an example of the roadblocks encountered). \ All of these algorithms are based on the Quantum Fourier Transform (QFT), on some version of what was introduced by Shor.

And then, of course, there's the simulation of quantum physics and chemistry. \ This was not only Feynman's original vision, but it's probably still, even after all these years, our best shot (at least that we know) at an economically useful application of quantum computers.

Now there \emph{are} other examples of exponential quantum speedups in the non-oracle world. \ For example, if you have a knot whose Jones polynomial you'd like to additively estimate at a root of unity, you're in luck! \ Aharonov, Jones, and Landau \cite{ajl} showed that not only is that problem in $\mathsf{BQP}$, it's $\mathsf{BQP}$-complete, which means that it captures precisely the power of quantum computation. \ This result in some sense builds on the work of Witten \cite{witten:jones}, who related the Jones polynomial to topological quantum field theories (TQFTs), as well as Freedman, Kitaev, and Wang \cite{fkw}, who showed that TQFTs encode all of quantum computation.

And then there are some extremely special problems in machine learning, such as estimating Betti numbers \cite{lgz}---a problem in topological data analysis---where there seem to be exponential quantum speedups, mostly based on a quantum algorithm from 2008 called the Harrow-Hassidim-Lloyd or HHL algorithm \cite{hhl}, for learning some information about implicitly specified but gigantic systems of linear equations.

There was hope for some years that quantum computing would provide much broader exponential speedups for machine learning problems, which would've been a huge deal if it were true. \ Unfortunately, many of those hopes have been killed, in part by a breakthrough four years ago by Ewin Tang \cite{tang}, who was then an 18-year-old undergraduate working with me in Austin, and then subsequent works building on hers, which have shown how to ``dequantize'' many of the quantum machine learning algorithms that we knew. \ That is, people found classical algorithms that were only polynomially slower than the quantum ones. \ So, it's a major challenge right now to find quantum machine learning speedups that are exponential and that matter in practice and that resist Ewinization.

People often ask, is that it? \ Well, it's not \emph{quite} it, but yes: the list of exponential speedups is smaller and more specialized than many people would like. \ People often say, we need more quantum algorithm techniques, beyond the Fourier transform. \ And that's true, but I would emphasize something different. \ We also need more targets for the quantum algorithms to hit. \ If there \emph{were} a fast quantum algorithm for $\mathsf{NP}$-complete problems, that could change civilization. \ But if we take that off the table---if we say that that's too unlikely---then a lot of what remains is kind of esoteric. \ So yes, give us more targets!

\section{The Yamakawa-Zhandry Breakthrough}

I am delighted to report that there \emph{was} a breakthrough, just this year, in finding an entirely new exponential quantum speedup. \ Takashi Yamakawa and Mark Zhandry \cite{yz} looked at the following task: you're given as input a program to compute a random or pseudorandom Boolean function, say $f:\{0,1\}^n\rightarrow \{0,1\}$, mapping $n$ input bits to $1$ output bit. \ You would like to find a list of $n$-bit strings, call them $x_1,\ldots,x_n$, each of which is mapped to $0$ by $f$:

$$ f(x_1) = \cdots = f(x_n) = 0.$$

So far that's easy: half of all strings are mapped to $0$. \ But we want the strings to have the additional property that, when we concatenate them all, we get an $n^2$-bit string which is a codeword of a particular error-correcting code that we're told.

This is not an especially useful task, you might say, but here is what Yamakawa and Zhandry proved about it:

\begin{enumerate}
\item There is a polynomial-time quantum algorithm to solve this problem, for an appropriate choice of the error-correcting code.

\item On the other hand, if $f$ is a truly random function, then any classical algorithm would need exponentially many queries to $f$ to solve the problem.
\end{enumerate}

Notice furthermore that, once you've found a solution, you can easily check it with your classical computer. \ So, the significance is that this is the first exponential quantum speedup for an $\mathsf{NP}$ search problem that's based on a \emph{random} or \emph{unstructured} black-box function. \ Because $f$ is random, the reigning ideology of cryptographers holds that you could replace it by a pseudorandom function, and the problem will almost certainly remain classically hard. \ But if $f$ is pseudorandom, then you could put this whole thing on a real quantum computer and run it---just like Shor's algorithm!

Unfortunately, this still won't work on a \emph{near-term} quantum computer. \ For, although the function $f$ is extremely simple, Yamakawa and Zhandry's algorithm is at least as complicated as Shor's algorithm. \ In particular, it requires you to run decoding of an error-correcting code on a coherent superposition of inputs. \ We expect algorithms of that sort to require full fault-tolerant quantum computers.

\section{Near-Term Exponential Speedups: Sampling-Based Quantum Supremacy}

This naturally raises the question: then what exponential speedups \emph{can} we demonstrate on a near-term device, without error-corrected qubits? \ So far, about all we've been able to think of has been the so-called \emph{sampling-based quantum supremacy experiments} (see, e.g., \cite{aark,achen,bjs}). \ These were just a theoretical idea a decade ago, but as you've heard, teams at Google, USTC, and Xanadu have actually carried these experiments out over the past three years.

One example of this sort of experiment is BosonSampling \cite{aark}, which Alex Arkhipov and I proposed in 2011, and which basically just involves sending a bunch of photons through a random network of beamsplitters and then measuring where they end up. \ In the analysis of BosonSampling, we used the fact that bosonic amplitudes are permanents of matrices, and the permanent function is what's called $\mathsf{\#P}$-complete, which is even above $\mathsf{NP}$-complete. \ Using this, we argued that a classical computer would almost certainly have a hard time sampling from the same probability distribution over lists of photon occupation numbers, given as input a description of the beamsplitter network.

BosonSampling (in a variation called Gaussian BosonSampling) was experimentally demonstrated last year, with about $100$ photons, by the group at USTC in China \cite{ustc}. \ I should say that, because of the noise in their system, it's debatable whether or in what sense they achieved quantum supremacy. \ Xanadu has since done a different BosonSampling experiment \cite{xanadu}, where the case for quantum supremacy might be stronger, for example because of their use of number-resolving detectors (which count the number of photons in each output mode, rather than only reporting whether the number is zero or nonzero).

\begin{figure}
\includegraphics[width=6.5in]{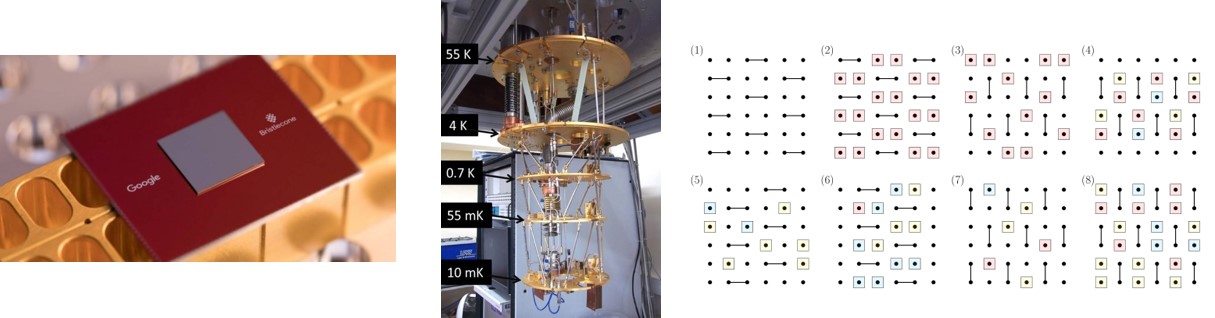}
\caption{Google's $53$-qubit Sycamore chip; the dilution refrigerator in which the chip is placed; a staggered pattern of nearest-neighbor gates applied to the qubits in such a chip.}
\label{google}
\end{figure}

A second example this sort of experiment is Random Circuit Sampling. \ This is what was used in the \emph{first} demonstration of quantum supremacy, by the group at Google in 2019 \cite{arute} (see Figure \ref{google}). \ The Google group said, let's take something like BosonSampling, but modified so that it naturally runs on a superconducting qubit system like ours. \ So, they had $n=53$ qubits in a $2$-dimensional lattice on a chip, called Sycamore, which was placed in a dilution refrigerator and cooled to about 10 milliKelvin (see Figure \ref{google}).

They initialized all the qubits to the $|0\rangle$ state, and then applied to them a random sequence of gates---i.e., a random circuit $C$---of depth $20$. \ Then they measured each of the qubits in the $\{ |0\rangle, |1\rangle \}$ basis. \ This produces a $53$-bit string as output.

By running the same random circuit $C$ over and over, with the qubits re-initialized to $|0\rangle$ each time, you get a list of $53$-bit strings---a bunch of independent samples $s_1,\ldots,s_k$ from a probability distribution, which we can call $D_C$. \ Crucially, $D_C$ is \emph{not} particularly close to the uniform distribution over $\{0,1\}^{53}$. \ Instead, because of random constructive and destructive interference, some output strings have higher probabilities than others, and this produces a statistical signal that can be checked.

Thus, the Google group validated the outputs $s_1,\ldots,s_k$ by calculating their ideal probabilities $p_{s_1},\ldots,p_{s_k}$ using a large classical computer, and checking that they were often larger than the average value of $2^{-n}$, consistently with what you'd predict from quantum mechanics and from knowledge of the circuit $C$. \ In particular, Google reported

$$ p_{s_1} + \cdots + p_{s_k} \ge \frac{bk}{2^n}$$

\noindent for $b\approx 1.002$. \ For comparison, classical random guessing would have yielded $b\approx 1$, while an ideal noiseless quantum computer would have yielded $b\approx 2$.

More broadly, switching attention from search problems or decision problems to these sampling problems has two big advantages. \ Mostly obviously, the sampling problems make it vastly easier to see an exponential speedup using currently existing quantum computers. \ That's because they involve operations like ``apply a random sequence of gates,'' or ``apply a random network of beamsplitters,'' rather than anything more finely sculpted.

Secondly, at least assuming an ideal quantum computer, exponential speedups over classical computers turn out to \emph{provable} for sampling problems---not unconditionally, but based on extremely secure complexity assumptions, like the non-collapse of the polynomial hierarchy, if you know or care what that means.

Having said that, a big disadvantage of the sampling tasks is that even just to verify the results using a classical computer seems to take exponential time. \ This means you can just barely do it with $50$ qubits---but if you went to $100$ or $200$ qubits, then if you sampled perfectly, you could never convince a skeptic of that, since no supercomputer on earth can do $2^{200}$ steps.

I should mention that there are some amazing recent protocols for verifying quantum computation---i.e., having a quantum computer cryptographically prove its answer to a classical skeptic, via the QC and the skeptic sending messages back and forth. \ Most notably, a breakthrough four years ago by Urmila Mahadev \cite{mahadev} showed how to make \emph{any} quantum computation verifiable in this way, assuming only the security of lattice-based encryption against quantum attack. \ Unfortunately, we don't yet know how to implement any of these interactive protocols using a near-term device---a familiar refrain! \ (Though see \cite{kcvy} for some steps toward a practical interactive protocol for demonstrating quantum advantage.)

A second disadvantage of the second tasks is that it's unclear whether they have any applications, beyond just proving quantum supremacy itself. \ I had a proposal a few years ago, that you could use these sampling tasks to generate cryptographically certified random bits, to distribute over the Internet---for proof-of-stake cryptocurrencies, for example. \ Unfortunately, you're still up against the exponential cost of verification. \ So to make that scheme practically useful, we would really want a fast way to classically verify the results.

\section{Quantum Speedups in the Black-Box Model}

Switching attention to the black-box model, I want to tell you a little bit about the exponential quantum speedups that we know there.

Famously, from near the very beginning of the field, there was Simon's problem \cite{simon}, which is an oracle problem, where you're given a black-box function $f:\{0,1\}^n \rightarrow \{0,1\}^n$, mapping $n$-bit strings to $n$-bit strings, and you're promised that $f$ encodes a secret called $s$, so that if I add $s$ bitwise to any input string, then $f$'s output is unaffected:

$$f(x) = f(x\oplus s)$$

\noindent Furthermore, $s$ is the only nonzero such string. \ Your challenge is to find $s$. \ It can be shown that any classical algorithm needs ~$2^{n/2}$ queries to $f$ to solve this problem, whereas Simon's quantum algorithm solves it using only $O(n)$ queries as well as $\operatorname{poly}(n)$ computation.

By the way, we still don't know a concrete, non-oracle instantiation of Simon's problem that wouldn't just be easy to solve using a classical computer. \ I continue to regard that as a beautiful open question.

Another example was a problem that I introduced in 2009, which I called Forrelation \cite{aar:ph}. \ Here you're given black boxes for \emph{two} Boolean functions, call them $f,g: \{0,1\}^n \rightarrow \{1,-1\}$, which map $n$ input bits to $1$ output bit, which it's convenient to treat as $1$ or $-1$. \ Individually, $f$ and $g$ are just uniformly random Boolean functions. \ The task is to decide whether $g$ is actually close to the Fourier transform of $f$, or whether the two functions are independent, promised that one of those is the case. \ The Fourier transform here means the Boolean Fourier transform,

$$\hat{f}(y) := \frac{1}{2^{n/2}} \sum_{x \in \{0,1\}^n} (-1)^{x\cdot y}f(x),$$

\noindent the thing that you get by applying Hadamard gates to each of $n$ qubits.

Andris Ambainis and I \cite{aa:for} showed that this problem actually gives a maximal separation between quantum and classical complexities in the black-box model. \ It's solvable using only $1$ query with a quantum computer, but it requires about $2^{n/2}$ queries by any classical algorithm. \ And it can be shown that every problem solvable with $1$ quantum query is solvable using $O(2^{n/2})$ classical queries.

The reason I introduced this problem was to try to prove a black-box separation between $\mathsf{BQP}$ and the Polynomial Hierarchy $\mathsf{PH}$. \ I was unable to do that, but just four years ago, Ran Raz and Avishay Tal \cite{raztal} managed to do it, proving that a variant of Forrelation is not in $\mathsf{PH}$ as a black-box problem. \ This was one of the great breakthroughs of quantum complexity theory of the past decade, along with $\mathsf{MIP*} = \mathsf{RE}$ \cite{mipre} for example.

\begin{figure}
\centering
\includegraphics[width=4in]{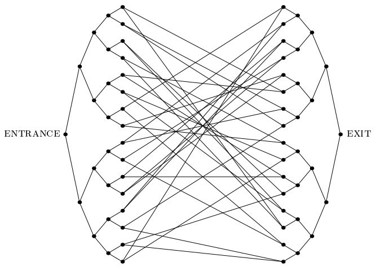}
\caption{The glued-trees graph, consisting of two complete binary trees connected at the leaves by a random cycle.}
\label{gluedtrees}
\end{figure}

We do know some other very intriguing exponential quantum speedups in the black-box model. \ One of them, from Childs et al.\ \cite{ccdfgs} in 2002, is a quantum walk on a certain very carefully designed graph. \ So, you have an ENTRANCE vertex on the left and an EXIT vertex on the right. \ You start at the ENTRANCE vertex, and your goal is to find the EXIT vertex. \ Connected to the ENTRANCE and EXIT are two complete binary trees, both of depth $n$, and then the trees' leaves are connected to each other by a random cycle (see Figure \ref{gluedtrees}).

What you can show is that, supposing that this graph is encoded by an oracle---meaning, given the label of any vertex, the oracle will tell you the labels of the neighbors of that vertex, but nothing else---if you classically start at the ENTRANCE and then walk on this graph, by any strategy you like, it will take you $\exp(n)$ time (to be precise, about $2^{n/2}$ time) to find the EXIT vertex. \ Whereas if you take a quantum walk on this graph, it will reach the EXIT vertex in expected polynomial time.

More recently, Matt Hastings \cite{hastings:adiabatic} achieved a breakthrough in 2020, when he was able to show that the adiabatic algorithm can get a superpolynomial speedup over any classical algorithm, for a very, \emph{very} carefully designed, weird fitness landscape. \ That was then improved by Gily\'{e}n and Vazirani (see \cite{gilyenvazirani}) to an exponential speedup.

Unfortunately, we don't yet know how to realize these by any ``real-world,'' non-oracle optimization problems, let alone any that would be useful in practice. \ That's another big challenge.

\section{The Need for Structure in Quantum Speedups: Lessons from the Black-Box Model}

In the black-box model, we also know some fundamental limitations on the possible quantum speedups that you can get. \ This started in 1994, when Bennett, Bernstein, Brassard, and Vazirani (BBBV) \cite{bbbv} showed that searching an unstructured list of size $N$ requires at least about $\sqrt N$ queries on a quantum computer. \ In other words, \emph{Grover's algorithm is optimal}---although at that time, Grover's algorithm hadn't even been discovered yet.

For many other ``unstructured'' problems---for example, computing the PARITY or the MAJORITY of $N$ input bits---there is no asymptotic quantum speedup at all. \ The proofs of such statements use the linearity of quantum mechanics combined with a quantum algorithm's inability to ``notice'' a small change well, supposing someone flipped just one bit of the oracle's output---unless the algorithm had presciently assigned a lot of amplitude to that one bit.

There's a beautiful generalization, due to Beals et al.\ \cite{bbcmw} from 1998, which says that for \emph{every} Boolean function $f$, the deterministic query complexity and the quantum query complexity are polynomially related to one another. \ They can differ by at most a power of $6$.  That was recently improved to a power of $4$ \cite{abkrt}, which we now also know is tight \cite{abblss}. \ For \emph{randomized} versus quantum query complexity, we still don't know exactly, but we've learned that the exponent of the largest possible separation is between $3$ and $4$ \cite{bansalsinha,sherstovsw}.

Grover's algorithm, of course, gives a power-$2$ (i.e., square-root) separation, for the $N$-bit $\operatorname{OR}$ function.

This result of Beals et al.\ explains why the exponential speedups from Simon's problem, or from period-finding, which is the heart of Shor's algorithm, require these very structured oracles. \ You need to be promised that your function \emph{is} periodic, or that it \emph{does} satisfy the Simon property. \ If you didn't have that, then no exponential quantum speedup would have been possible.

One of the main things that I studied as a graduate student was a sort of intermediate case called the \emph{collision problem}. \ Here you're given a 2-to-1 function $f:\{1,\ldots,N\}\rightarrow \{1,\ldots,M\}$ ($M\ge N$), and you just want to find two inputs $x,y$ with the same output, $f(x)=f(y)$. \ By the birthday paradox, with a classical computer, about $\sqrt{N}$ queries to $f$ are necessary and sufficient to find one of these collisions. \ Brassard, H{\o}yer, and Tapp \cite{bht} noticed that one can cleverly combine the birthday paradox with Grover's algorithm to get an $N^{1/3}$-time quantum algorithm for finding collisions.

So then the question arose, could you do even better than that with a quantum computer? \ What's striking is that a quantum computer, in some sense, can \emph{almost} find a collision using only $1$ query. \ To do so, you just make an equal superposition over all inputs $x$, query $f$, and then measure the $f(x)$ register:

$$ \frac{1}{\sqrt{N}} \sum_{x=1}^{N} |x\rangle \rightarrow \frac{1}{\sqrt{N}} \sum_{x=1}^{N} |x\rangle |f(x)\rangle \rightarrow \frac{|x\rangle+|y\rangle}{\sqrt{2}} |f(x)\rangle$$

\noindent where $f(y)=f(x)$.

You then have the collision pair right in your hand, as it were: an equal superposition over two inputs $x$ and $y$ with the same $f$-value. \ If only you could take this quantum state and measure it twice, you'd be done.

Nevertheless, what Yaoyun Shi and I showed in 2002 \cite{as} is that alas, the Brassard-H{\o}yer-Tapp algorithm is optimal. \ We of course use the fact that, in actual quantum mechanics, you \emph{can't} measure multiple times!

Later, Andris Ambainis and I \cite{aa:struc} generalized the collision lower bound to \emph{any} partial function that's symmetric under permuting the inputs and outputs, like the collision function is. \ For any such function, we showed that quantum computers can give you at most a $7^{th}$-power speedup over classical ones. \ That was recently improved by Chailloux \cite{chailloux} to show that it's most a $3^{rd}$-power speedup---and moreover, even for functions that are symmetric under permuting the inputs only. \ Ben-David et al.\ \cite{bendavid:symmetry} recently had a beautiful result generalizing the conclusion to yet other symmetries. \ For example, for any partial function of a graph adjacency matrix---anything that's invariant under permuting the graph---quantum computers give you at most a polynomial speedup over classical ones in terms of query complexity.

Personally, I like to think about black-box problems in terms of a ``hierarchy of structure,'' where the more structure you impose on a black-box function $f$, the greater the chance that there's an exponential quantum speedup for learning properties of $f$.

\begin{figure}
\centering
\includegraphics[width=4in]{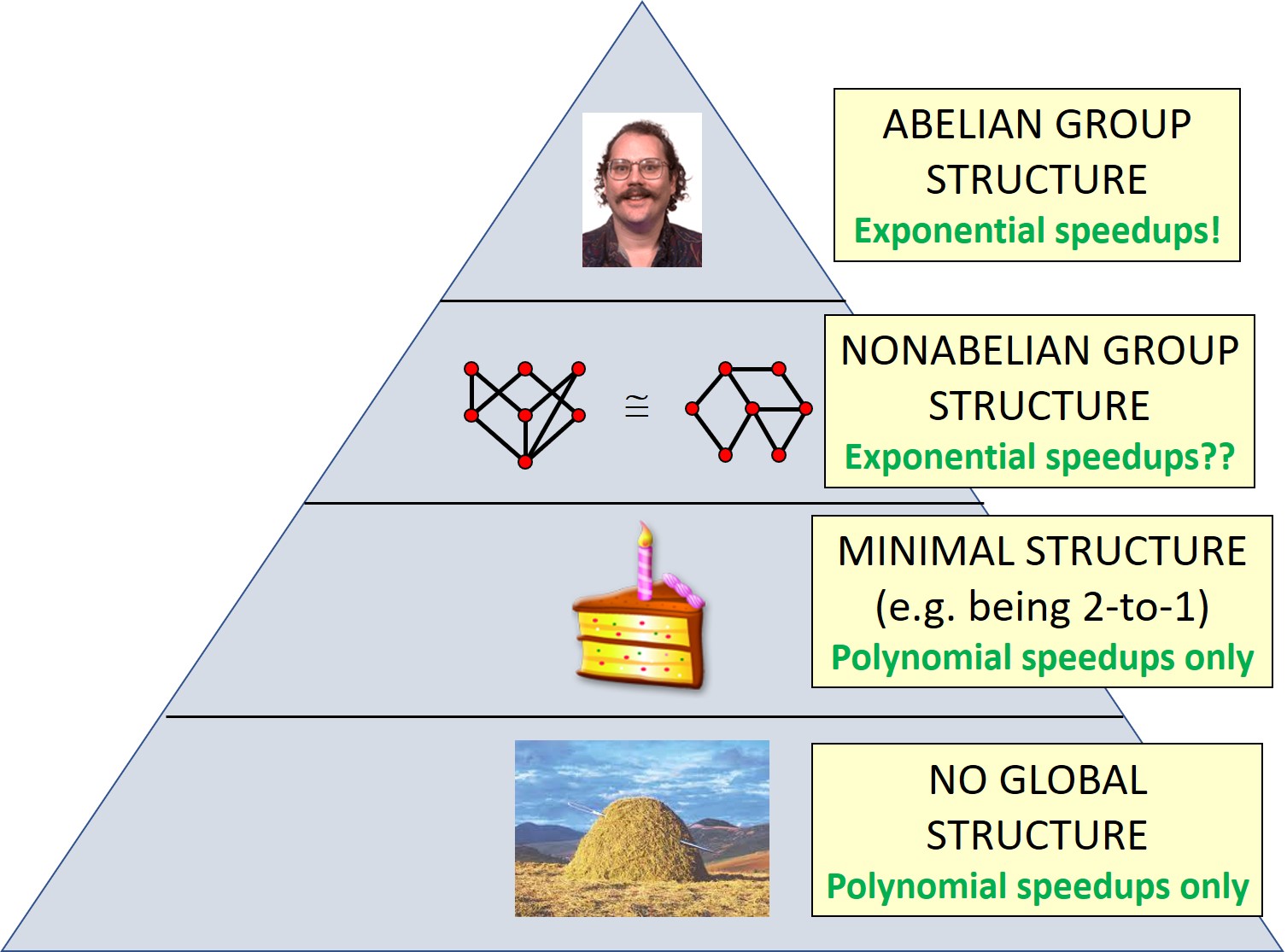}
\caption{Part of the ``hierarchy of structure.''}
\label{hierarchy}
\end{figure}

Of course, the full ``hierarchy'' defies any simple description: there are infinitely many possible kinds of structure, we expect only a special subset of them to allow exponential quantum speedup, and it seems too much to hope for a full characterization of which types of structure suffice---just like it would be too much to hope for a full characterization of which problems admit efficient \emph{classical} algorithms. \ Having said that, one large chunk of what quantum algorithms researchers have cared about \emph{does} organize itself into a relatively simple four-part hierarchy (see Figure \ref{hierarchy}).

At the bottom, we have problems with no global structure: just pure finding a needle in a haystack (or, say, computing the PARITY or MAJORITY of $N$ input bits, with no promise on those bits). \ Here we've seen that quantum computers can give polynomial speedups only.

Then we have problems with minimal structure, like the function being 2-to-1. \ There again we have polynomial speedups only, although for subtler reasons---reasons that would fail if (for example) we could make multiple non-collapsing measurements.

Then we have problems with a \emph{non}-abelian group structure. \ A famous example here would be the graph isomorphism problem, which I didn't talk about earlier, but which has now been a target (so far, an unsuccessful target) of quantum algorithms research for 27 years. \ We don't know yet whether graph isomorphism or related isomorphism problems are in $\mathsf{BQP}$. \ We do know that, in some sense, the barrier can't be one of query complexity. \ A 1997 result of Ettinger, H{\o}yer, and Knill \cite{ehk} shows that the ``nonabelian hidden subgroup problem,'' which includes graph isomorphism as a special case, has polynomial quantum query complexity. \ The ``only'' problem is that the quantum circuits that extract the final answer given the query results might be exponentially large.

Then, at the top, we have problems with abelian group structure. \ As I said, problems with that much structure typically \emph{do} admit exponential quantum speedups, because of Shor's algorithm and its variants.

Thinking about the role of structure in black-box quantum speedups, Ambainis and I \cite{aa:struc} conjectured a generalization of the result of Beals et al.\ \cite{bbcmw}, that no superpolynomial speedups are possible for total Boolean functions. \ Our generalization says that for \emph{any} quantum algorithm that makes $T$ queries to an oracle function $f$, there exists a classical algorithm that makes only $\operatorname{poly}(T)$ queries to $f$, and that approximates the quantum algorithm's acceptance probability on \emph{most} values of the oracle---i.e., with high probability over a random choice of $f$.

We showed that this would follow from an extremely natural conjecture about influences in low-degree polynomials---a conjecture whose statement makes no reference to quantum computing. \ Despite significant effort by experts, this ``Aaronson-Ambainis Conjecture'' remains open a decade later (though see \cite{bsw} for exciting recent progress).

In posing our conjecture, my and Ambainis's goal was to formalize the intuition that, relative to a \emph{random} oracle, the only quantum speedup you should expect is the kind from Grover's algorithm, the polynomial kind. \ Exponential speedups should require structured oracles.

You might wonder, what about the breakthrough by Yamakawa and Zhandry \cite{yz} that I mentioned before? \ Didn't they just refute our conjecture?

Interestingly, they didn't. \ What they did instead was to \emph{evade} it! \ They evaded our conjecture by looking not at decision problems, but at \emph{$\mathsf{NP}$ search problems}, which Ambainis and I hadn't talked about. \ For those, it turns out that you \emph{can} get exponential quantum speedups on random oracles.

\section{Future Directions and Conclusions}

In my view, an extremely urgent problem is to make near-term quantum supremacy verifiable. \ Can we devise an interactive protocol where

\begin{enumerate}
\item[(1)] a challenger creates a pseudorandom quantum circuit that conceals some secret,
\item[(2)] the challenger sends the circuit to a quantum computer, and then
\item[(3)] the quantum computer has to find the secret by running the circuit?
\end{enumerate}

\noindent We don't yet know how to implement this idea. \ We have some limited no-go theorems, for example with BosonSampling \cite{anguyen,berkowitzdevlin}, but I'm still optimistic that this will turn out to be possible.

Here's a concrete question that as far as I know hasn't been asked before. \ Let $C$ be a quantum circuit on $n$ qubits and, let's say, $n^2$ gates. \ Suppose that $C$ is chosen uniformly at random from among all such circuits that output some one basis state, call it $|x\rangle$, with a large probability, say at least $0.1$. \ What does a random quantum circuit \emph{of that kind} look like? \ If I handed you such a circuit, would it be classically intractable to determine the particular string $x$ that was output with high probability?

Why do I care? \ Well, suppose the above problem \emph{was} hard. \ Then if only we could efficiently \emph{generate} these circuits (in a way where we knew $x$), they would be a feasible route to near-term quantum supremacy, of the kind where a classical computer could efficiently verify the answer. \ On the other hand, even if finding pseudorandom peaked circuits remained hard, I'd still love to know whether those circuits \emph{exist}.

Let me conclude with a vague question. \ When it comes to exponential quantum speedups, is there a ``Law of Conservation of Weirdness''?  I.e., for every problem that admits an exponential quantum speedup, must there be some weirdness in its detailed statement, which the quantum algorithm exploits to focus amplitudes onto the rare right answers? \ I would love for this not to be true! \ But I think that for thirty years, it's been the conjecture to beat, and we're still trying. \ Thank you.

\section{Acknowledgments}

I'm grateful to Umesh Vazirani for inviting me to give this talk, to the participants in the Solvay Conference for their feedback, and to Alexandre Sevrin for pestering me to produce this writeup.

\bibliographystyle{plain}
\bibliography{thesis}

\end{document}